# Ultrafast Spin Accumulations Drive Magnetization Reversal in Multilayers

Harjinder Singh[1], Alberto Anadón[1], Junta Igarashi[2], Quentin Remy[3], Stéphane Mangin[1], Michel Hehn[1], Jon Gorchon[1,*], Gregory Malinowski[1,†]

[1]*Université de Lorraine, CNRS, IJL, Nancy, F-54000, France*
[2]*National Institute of Advanced Industrial Science and Technology, Tsukuba 305-8563, Japan*
[3]*Department of Physics, Freie Universität Berlin, 14195 Berlin, Germany*

Engineering and controlling heat and spin transport on the femtosecond time-scale in spintronic devices opens up new ways to manipulate magnetization with unprecedented speed. Yet the underlying reversal mechanisms remain poorly understood due to the challenges of probing ultrafast, non-equilibrium spin dynamics. In this study, we demonstrate that typical magneto-optical experiments can be leveraged to access the time evolution of the spin accumulation generated within a magnetic multilayer following an ultrafast laser excitation. Furthermore, our analysis shows that the final magnetic state of the free-layer in a spin-valve is mainly dictated by the ultrafast dynamics of the reference-layer magnetization. Our results disentangle magnetization and spin transport dynamics within a multilayer stack and identify demagnetization and remagnetization-driven spin accumulation as the key mechanism for all-optical switching. These findings establish new design principles for ultrafast spintronic devices based on tailored spin current engineering.

Since the discovery that the magnetization of magnetic materials can be manipulated by light [1,2], achieving an ultrafast reversal of the magnetization via laser pulses has become a central goal in spintronics [3]. All-optical switching of magnetization was first demonstrated in GdFeCo ferrimagnetic alloys [4] where the reversal was attributed to a local ultrafast transfer of angular momentum between the two antiferromagnetically coupled sublattices [4,5]. In multilayers composed of two magnetic layers separated by a metallic spacer, also known as spin-valves structures, the demagnetization of one of the layers was shown to influence the magnetizations dynamics of the other [6,7]. This non-local influence has been attributed to spin-currents generated during the demagnetization process which can be injected into the second layer [8–11]. If such spin currents inject enough opposite angular momentum and heat, the magnetization of the layer can be even reversed [12–15]. Recently, such reversal was observed for the first time in fully ferromagnetic spin-valves [15]. Surprisingly, the results showed that optical excitation of the system in its parallel (P) configuration (i.e. parallel magnetizations between layers) could promote a switching into its antiparallel (AP) alignment, hinting at some unknown source of opposite angular momentum in the system. One possible mechanism considers that one of the layers will have a strong re-magnetization dynamics. The re-absorption of majority spins during cooling leads to an injection of minority spins into the second layer, which could result in switching [15]. However, the experimental reversal timing appeared to occur before the remagnetization of the reference layer, requiring another mechanism to be invoked. Igarashi et al. [15] suggested that the spin current generated by one layer during its demagnetization could suffer scattering at the interface between the spacer and the second layer, resulting in a back-propagating minority spin current. This effect is similar to what has been reported in the context of spin-transfer torque (STT) experiments [16]. Using an extension of existing s-d models [17] that included reflections with spin flip, Remy et al. [18] reproduced the experimental results. Nevertheless, no clear experimental proofs have been brought forward to settle this open question, partially due to the challenge in detecting such spin currents within deep layers. While spin accumulation has been optically observed at the surface of non-magnetic metals [19–21] recent work shows that such accumulations may also be accessible through ferromagnetic layers [22].

In this study, we demonstrate that the time-resolved magneto-optical Kerr signals measured in magnetic spin-valves result from a superposition of local magnetization dynamics and the spatio-temporal evolution of the spin accumulation within the multilayer. Our analysis supported by simulations based on an extended s-d model [18] unambiguously reveals that the reversal of the switching (free) layer in the spin-valve structure results from the spin current generated by the demagnetization (resp. re-magnetization) of the reference layer if we start from an antiparallel (resp. parallel) alignment. By combining rotation and ellipticity measurements, we further extract the time-resolved spin accumulation originating from both ferromagnetic layers. These results are in good agreement with a spin-diffusion model [20,23] with a single free parameter.

The studied samples were grown on a glass substrate by magnetron sputtering. The multilayers structure is Ta(5)/Pt(4)/[Co(1)/Pt(1)]3/Co(1)/Cu(80)/[Co(0.6)/Pt(1)]2/Ta(5) and is identical to the spin valve structure used by Igarashi et al. [15] where the numbers in parenthesis represent the thickness in nm. Both ferromagnetic layers exhibit strong perpendicular magnetic anisotropy (see Supp. Mat.) [24]. As shown in Fig.1a, we refer to the top [Co(0.6)/Pt(1)]2 layer as the *free* layer, whereas the bottom [Co(1)/Pt(1)]3/Co(1) layer is named the *ref* layer. The distinct coercivities allow us to stabilize both parallel (P) and antiparallel (AP) magnetization configurations. The 80 nm thick Cu spacer was chosen to ensure that the probe only senses a single ferromagnetic layer, further simplifying the analysis, while the pump is always incident on the *free* layer (see Supp. Mat.) [24]. In this sample, the P-to-AP reversal of the *free* layer magnetization was confirmed using single-shot and time-resolved measurements [15]. However, we are revisiting the time-resolved measurements due to our recent advancements in experimental and theoretical approaches. For this, a Ti: sapphire fs laser is used, which delivers pulses of approximately 25 femtoseconds (fs) at a 5 kHz repetition rate (see Supp. Mat.) [24]. We conduct pump-probe time-resolved magneto-optical Kerr effect (TR-MOKE) measurements to extract the dynamics of the *free* and *ref* layers. We measure both the Kerr rotation ($\theta_k$) and ellipticity ($\epsilon_k$) as a function of the time delay between the pump and probe, and derive a normalized pump-induced Kerr rotation ($\Delta\bar{\theta}_K$) and ellipticity ($\Delta\bar{\varepsilon}_K$) (see Supp. Mat.) [24]. We remind that a basic assumption in time-resolved magneto-optics is that, at all times, the relation $\theta_k \propto \epsilon_k \propto M$ should remain valid, with $M$ being the magnetization. In other words, as long as the probe is only sensing $M$, we should have $\Delta\bar{\theta}_K = \Delta\bar{\varepsilon}_K$.

---

[*] jon.gorchon@univ-lorraine.fr
[†] gregory.malinowski@univ-lorraine.fr

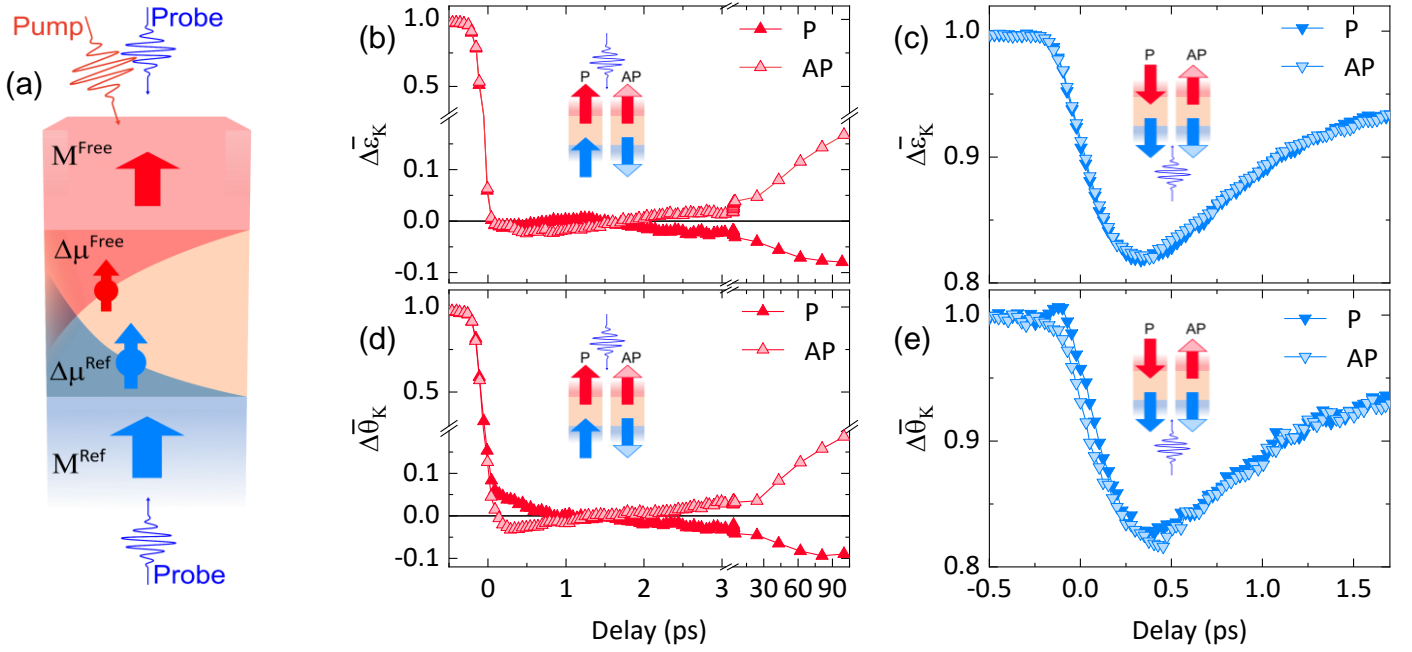

Fig. 1 Schematic representation and time-resolved magneto-optical Kerr effect (TR-MOKE) measurements in polar geometry. (a) An illustration of the sample structure with the pump and probe (the pump is actually at normal incidence, see Supp. Mat)[24]. The top layer is the [Co/Pt]$_2$ free layer, and the bottom layer is the [Co/Pt]$_3$ ref layer, which are separated by the Cu spacer. (b,d) The probe is incident on the free layer (red triangles) or (c,e) on the ref layer (blue triangles). For both layers, we plot either (b,c) the Kerr ellipticity ($\varepsilon_K$) or (d,e) the Kerr rotation ($\theta_K$) in both P and AP configurations. The P or AP state is set by an external magnetic field.

*TR-MOKE dynamics of free and reference layers- We* first examine the measured $\Delta\bar{\varepsilon}_K$ as shown in Fig. 1b and c. When probing the *free* layer and starting from a P state (dark red triangles), we observe a clear switching to the AP state at around 1.5 ps. However, the switching is not observed when starting from the AP state (light red triangles) as the signal returns to its initial value, in agreement with Ref. [15]. We also note a subtle difference between the P and AP dynamics before 1.5 ps, which we will address later. We then measure $\Delta\bar{\varepsilon}_K$ on the *ref* layer while the pump is on the *free* layer (Fig.

1c). There, we observe no difference in the dynamics between the P and AP configurations, both demagnetizing by ~18 % and peaking at ~0.3 ps. We then conduct the same experiments while measuring $\Delta\bar{\theta}_K$. As shown in Fig. 1d, the measured dynamics on the *free* layer exhibit similarities with $\Delta\bar{\epsilon}_K$ over the longer timescales: a reversal when exciting the P state (dark red triangles) and no reversal when exciting the AP state (light red triangles). However, striking differences appear during the first hundreds of femtoseconds, with a slowdown of $\Delta\bar{\theta}_K$ when starting in the P state, versus an acceleration and reversal of sign otherwise. As these short time-scale features were not observed by Igarashi et al. [15] we double-checked the dynamics under similar earlier conditions (300 fs pulses centered at 785 nm). Again, we obtain very similar results to the published data (see Supp. Mat.) [24]. We can thus infer that the change in probe wavelength and pulse duration are critical for the observed details in the dynamics. We postulate that these differences observed during the initial 1.5 ps stem from the fact that the probe not only senses the magnetization but also the spin accumulation in the multilayer structure, resulting from the ultrafast demagnetization of both layers. The sign and amplitude of the magneto-optical signal induced by the magnetization and spin accumulation are strongly wavelength dependent [25,26], so such differences in the dynamics are to be expected. When measuring $\Delta\bar{\theta}_K$ in the *ref* layer (blue triangles in Fig. 1e), we again observe a noticeable difference in the dynamics during the first hundreds of femtoseconds between the P and AP configurations. Due to the thick Cu spacer, the *ref* layer is only excited by the hot electron current resulting from the optical absorption in the *free* layer [27]. However, these hot electrons not only carry heat but also spin angular momentum. Therefore, we attribute the observed differences to the spin accumulation. In contrast, the absence of differences in Fig. 1c indicates that the $\epsilon_k$ has likely little to no sensitivity at all to the spin accumulation in the structure. The higher sensitivity of $\theta_k$ to accumulated spins is partially confirmed by measurements of spin accumulation on the surface of a thick Cu layer (see Supp. Mat.) [24], which echo recent findings in [Co/Ni]/Cu multilayers [22].

*Spin accumulation dynamics-* As described in Hamrle et al. [28] the total complex magneto-optical Kerr vector defined as $\Theta_k = \theta_k + i\epsilon_k$, can be decomposed into a sum of sub-vectors associated to each sub-magnetic layer (or subsystem). In our case, we consider $\Theta_k = \Theta_k^M + \Theta_k^{\Delta\mu_s,FM} + \Theta_k^{\Delta\mu_s,Cu}$, where $\Theta_k^M$ is associated to the probed Co/Pt magnetization, $\Theta_k^{\Delta\mu_s,FM}$ to the spin accumulation in the ferromagnet (FM) ($\Delta\mu_s^{FM}$) and $\Theta_k^{\Delta\mu_s,Cu}$ to the spin accumulation in the Cu layer ($\Delta\mu_s^{Cu}$) (see Supp. Mat.) [24]. Different strategies exist then to disentangle such static [28] and/or dynamic [29] signals. For example, in Refs. [29,30] a specific projection axis in the complex Kerr plane is selected via a quarter-waveplate, so that the projection of one of the sub-vectors is null and sensitivity to the other sub-vectors remains. However, this procedure requires prior knowledge of such specific complex angle. Generally, it is found by performing multiple hysteresis measurements as a function of the quarterwaveplate angle until an extinction of one of the contributions is found. In this work, similarly to Ref [22], we instead measure pure time-resolved rotation and ellipticity signals to find $\Delta\Theta_k$, as well as the corresponding static vector $\Theta_k^{M_s}$ (See Supp. Mat.). After normalization, we obtain $\Delta\bar{\Theta}_k = (\Delta\Theta_k^M + \Delta\Theta_k^{\Delta\mu_s,FM} + \Delta\Theta_k^{\Delta\mu_s,Cu})/\Theta_k^{M_s}$. We define $\Theta_k^M = (a + ib)M$, $\Theta_k^{\Delta\mu_s,FM} = (a' + ib')\mu_s^{FM}$ and $\Theta_k^{\Delta\mu_s,Cu} = (a'' + ib'')\mu_s^{Cu}$ where $a, a', a'', b, b'$ and $b''$ are the magneto-optical sensitivities. Taking the difference between the real ($\Delta\bar{\theta}_K$) and imaginary ($\Delta\bar{\varepsilon}_K$) parts of $\Theta_k$, the sensitivity to the magnetization is eliminated, and only sensitivity to spin accumulation $\Delta\mu_s$ is obtained:

$$\Delta\mu_s = \Delta\bar{\theta}_K - \Delta\bar{\varepsilon}_K = \frac{1}{M_s}\frac{b'}{a}\left(\frac{a'}{b'} - \frac{a}{b}\right)\Delta\mu_s^{FM} + \frac{1}{M_s}\frac{b''}{a}\left(\frac{a''}{b''} - \frac{a}{b}\right)\Delta\mu_s^{Cu} \quad (Eq.1)$$

In Fig. 2a we plot the spin accumulation signals $\Delta\mu_s$ in the P ($\Delta\mu^P$, filled circles) and AP ($\Delta\mu^{AP}$, open circles) configurations when probing the *free* layer, as derived from Eq.1. In order to isolate the spin accumulations induced by the *free* $\Delta\mu_s^{Free}$ and *ref* $\Delta\mu_s^{Ref}$ layers, as illustrated in (a), we will assume that the total spin accumulation $\Delta\mu_s$ is the result of their sum, $\Delta\mu_s = \Delta\mu_s^{Free} + \Delta\mu_s^{Ref}$. Therefore, by performing (half-)sums and differences between P and AP signals, we can isolate single contributions $\Delta\mu_s^{Free} = 0.5(\Delta\mu_s^P + \Delta\mu_s^{AP})$ shown in (b) and $\Delta\mu_s^{Ref} = 0.5(\Delta\mu_s^P - \Delta\mu_s^{AP})$ in (c). On the one hand, $\Delta\mu_s^{Free}$ results in a unipolar spin accumulation, as the *free* layer is strongly demagnetized, and the remagnetization is slow [31–33]. On the other hand, $\Delta\mu_s^{Ref}$ shows a bipolar character, typical of a strong remagnetization, as the one shown in Fig. 1c. We can notice that such remagnetization-associated negative spin-accumulation is potentially responsible for the reversal of the *free* layer, a scenario which we will revisit later. In (d) we plot $\Delta\mu^P$ and $\Delta\mu^{AP}$ as measured on the *ref* layer side, and in (e) and (f), we perform similar sum and difference operations to obtain $\Delta\mu_s^{Free}$ and $\Delta\mu_s^{Ref}$. The major differences between plots in (b) and (e), as well as (c) and (f), are the arrival times of the peaks. When probing $\Delta\mu_s^{Free}$ via the *free* layer (b), the peak of spin accumulation is immediate. However, when probing $\Delta\mu_s^{Free}$ through the *ref* layer, a delay for diffusion through the Cu is necessary (e). Once this *free* layer spin and energy reaches the *ref* layer via diffusion, it drives the demagnetization of the *ref* layer, which itself generates a new spin current. For this reason, $\Delta\mu_s^{Ref}$ as probed on the *ref* layer (f) has approximately the same delay as $\Delta\mu_s^{Free}$ probed on the *ref* layer (e). Finally, when probing $\Delta\mu_s^{Ref}$ via the *free* layer (c), the delay is maximal, since energy needs to diffuse down to the *ref* layer first, to generate a spin accumulation that can then diffuse back to the top of the stack.

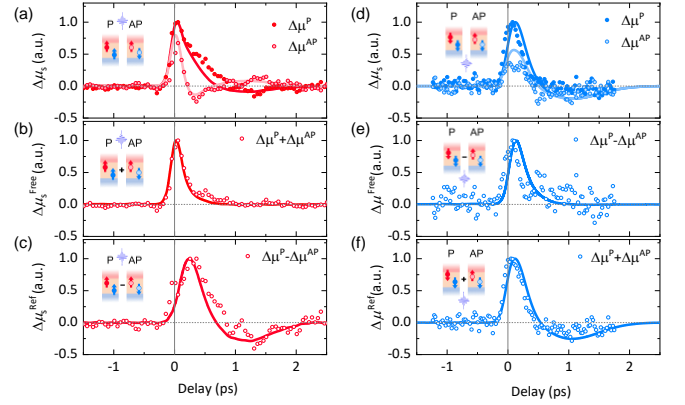

Fig. 2 Spin accumulation magneto-optical signals ($\Delta\mu_s$). We ascribe the quantity $\Delta\mu_s = \Delta\bar{\theta}_K - \Delta\bar{\varepsilon}_K$ to a spin accumulation. Data points in (a,d) are obtained from the difference between the normalized Kerr rotation and ellipticities. The top panels (a,d) show the mixed spin accumulation. The central panels (b,e) show the spin accumulation originating from the free layer's demagnetization, and is obtained by sum and difference operations from the top panel's data. The bottom panels (c,f) show the spin accumulation originating from the ref layer's demagnetization, and is similarly obtained by the complementary difference and sum operations. Solid lines are fits using the spin-diffusion model [21–23].

We perform numerical simulations using a spin-diffusion model [21–23] (see Supp. Mat.) [24], assuming a simplified Co/Cu/Co structure with only 8 independent parameters. We obtain the sources of spin in both *free* and *ref* layers by differentiation of their magnetization dynamics in Fig. 1b and (c). All parameters are fixed from the literature (see Supp. Mat.) except the *free* layer's Co/Cu interface spin conductance which is reduced by a factor of 3 with respect to the other interface, in order to fit the data. The resulting fits, shown as lines in Fig. 2 present extremely good agreement with the data. The delays, widths of the pulses as well as the bipolar/unipolar character are all captured. The reduced conductance of the *free* layer interface could be related to the growth of the Co on the thick Cu. If we plot instead the spin accumulation within the Co layers, the spin accumulation is dominated by each magnet's own spin generation and we cannot reproduce the data (see Supp. Mat.) [24]. Therefore, our simplified simulations suggest that in our experiments most of the spin accumulation signal would come from the Cu layer. The small discrepancies of the fit could be related to some spin-dependent transport at the interfaces and/or some remaining sensitivity to spin within the Co layers. These aspects require further exploration, particularly to be able to attribute the signals to the Cu and/or Co/Pt layers. A possible reason that could explain an enhanced sensitivity in the Cu layer is a potential similarity between the magneto-optical sensitivities within the magnet to both M and $\Delta\mu_s^{FM}$.

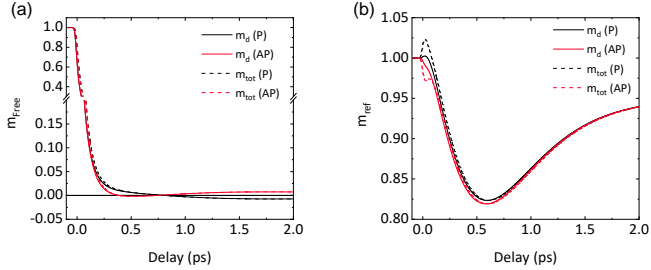

Fig. 3 Simulated magnetization dynamics using the s-d model. Normalized d electrons magnetization ($m_d$ - full line) and normalized total magnetization ($m_{tot}$ - dash line) of the (a) free and (b) reference layers starting from either a parallel (P - black) or antiparallel (AP - red) configuration. The laser fluence used for the simulation was 20 mJ/cm$^2$.

In other words, if the phase of $\Theta_k^M$ and $\Theta_k^{\Delta\mu_s,FM}$ (See Supp. Mat.) [24] are similar, i.e. $a'/b' \sim a/b$, then Eq.1 simplifies to $\Delta\mu_s \sim \Delta\mu_s^{Cu}$. To sum up, regardless of whether the signals are in the Co/Pt or spin accumulation within the few top and bottom nanometers.

To get a better insight into the mechanisms involved in the magnetization reversal and the observed magnetization dynamics, we modeled the ultrafast demagnetization and the spin current generation using an extended *s-d* model [17,18,34] (see Supp. Mat.) [24]. This model considers the coupling of localized *d* electrons with itinerant spin carried by the *s* electrons. For simplification we considered that the spin generated during the demagnetization of both magnetic layer propagates ballistically through the Cu layer, without scattering at the interfaces. Therefore, the spin generation and its polarity is directly related to the dM/dt. We can thus calculate the dynamics of the magnetization carried by the *d* electrons ($m_d$) as well as the dynamics of the total spin angular momentum carried by both the *s* (spin accumulation) and *d* electrons ($m_{tot}$). Fig. 3 shows the simulated magnetization dynamics for both the *free* (a) and *ref* layers (b). In the case of the free layer, the simulated dynamics show the same behaviour as experimentally observed when measuring $\Delta\theta_k$ (Fig. 1d). Indeed, the calculations reproduce the enhancement or reduction of the demagnetization during the first hundred of femtosecond when starting from the AP or P state, respectively. Moreover, the magnetization reversal is only obtained when starting from the P state. It occurs after 800 fs which is just after the reference layer starts remagnetizing (Fig. 3b). In the case of the reference layer, in agreement with the experimental results obtained when measuring $\Delta\theta_k$, we observe an increase (decrease) of the magnetization during the first hundred of femtosecond when starting from a P (AP) state which is induced by the large spin accumulation generated by the quasi-total demagnetization of the free layer. This difference between the P and AP configurations is large when considering the total magnetization, i.e. the sum of the *d* electrons magnetization and the spin accumulation, but less pronounced in the magnetization of the *d* spins.

The very good agreement between our set of simulations and experimental observations allow us to unambiguously conclude that the reversal of the free layer is induced by the spin accumulation generated by the remagnetization of the reference layer. We can also infer that the magneto-optical signal induced by the spin accumulation is mainly detectable in the Kerr rotation signal using our experimental conditions. These results complete the understanding of all-optical switching in ferromagnetic spin-valves: At low fluences, when the *ref* layer experiences a limited demagnetization, its remagnetization drives the P-to-AP reversal. However, at larger fluences, as the dynamics of the magnetization recovery of the ref layer slows down [31–33], remagnetization spin currents are reduced while the demagnetization spin currents are still large enabling the AP-to-P reversal [15]. Therefore, by engineering the demagnetization and remagnetization dynamics, one should be able to control the final state of the *free* layer.

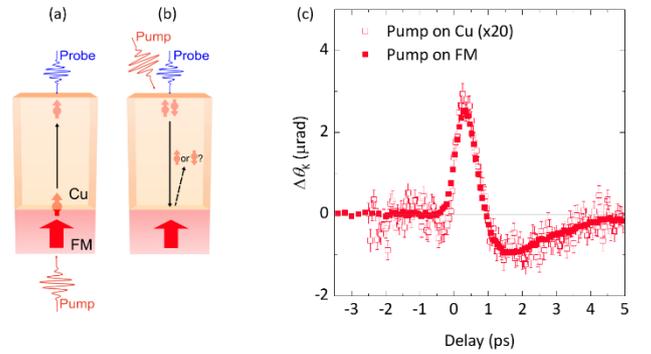

Fig. 4 Reflection mechanism- Schematic representation of spin accumulation generated when the pump (shown by red pulse) is on (a) the [Co/Pt]$_3$ ferromagnet (FM) or (b) Cu (100 nm) wheras the probe is always on the Cu side. (c) Kerr rotation in Cu as a function of time.

*Reflection mechanism-* We now turn to the STT-like mechanism proposed by Igarashi et al. [15] and contrast it with the interpretation presented in this work. In their scenario, it is assumed that the ultrafast demagnetization of the free layer emits a mixture of majority and minority spins. Upon reaching the Cu/reference-layer interface, minority spins could be preferentially reflected back toward the free layer, while majority spins are absorbed−resulting in a net angular momentum transfer that might account for magnetization reversal, even before the onset of remagnetization in the reference layer. However, a careful analysis shows that a mere 2% error in the estimation of the static magneto-optical signal $\Theta_k^{M_s}$ can significantly affect the switching time (see Supp. Mat.) [24], rendering the reversal compatible with a remagnetization mechanism. To experimentally test the presence of the STT-like mechanism we deposit a single [Co(0.8)/Pt(1)]$_3$ ferromagnet (FM) with a 100 nm Cu overlayer as shown in Fig. 4a,b and use an 80 MHz 300 fs Ti: Sapphire fs laser system (see Supp. Mat.) [24]. As shown in Fig. 4a, we first pump the FM and probe on the Cu side. In Fig. 4c (closed red squares) we plot the induced Kerr rotation, which we attribute to the demagnetization induced spin injection into the Cu layer [19,22]. We then pump the Cu as shown in Fig. 4b while the probe is still on the Cu. In this case, the small absorption of optical energy at the Cu surface induces a hot electron unpolarized current, i.e., a 50/50 mix of up and down spins which propagates until the FM layer. If the STT-like mechanism is dominant, an opposite spin accumulation should then be observed. However, as shown in Fig. 4c (open red squares), the spin accumulation has the exact same shape as that induced by a demagnetization of the FM layer, which is just due to the heating of the FM by the hot electron current. These findings provide strong evidence against the presence of an STT-like reflection mechanism in our system. We note that this experiment does not directly test the possibility of spin-flip reflection at the interface, as proposed in Ref. [18]. Nonetheless, the excellent agreement between our simulations and experimental data in Fig. 2 also shows no indication of such reflected spin components, further reinforcing our conclusion that the magnetization reversal in these spin-valves is governed by spin currents generated during remagnetization, rather than reflection-driven processes.

In summary, we have provided the first direct experimental disentanglement of ultrafast spin accumulation and magnetization dynamics in ferromagnetic spin-valves undergoing all-optical switching. By combining time-resolved Kerr rotation and ellipticity measurements with spin-diffusion and s-d model simulations, we resolve spin accumulation contributions originating separately from each magnetic layer. This decomposition reveals distinct unipolar and bipolar spin accumulation profiles, respectively associated with the demagnetization of the free layer and the remagnetization of the reference layer-unveiling a previously inaccessible temporal structure of spin transport in these heterostructures.

Our results show that the reversal of the free-layer magnetization in the spin-valve from the parallel to the antiparallel state is driven by remagnetization-generated spin currents from the reference layer, a mechanism that is quantitatively captured by both our spin-diffusion and s-d models. Importantly, we find no experimental evidence for the presence of STT-like backscattered spin currents, as confirmed by control experiments in Cu/FM heterostructures, further reinforcing the central role of remagnetization-driven spin transport. Importantly, beyond elucidating the ultrafast, single-pulse switching mechanism in ferromagnetic spin-valves, our work establishes a robust magneto-optical framework for probing femtosecond spin accumulation dynamics in a broad range of multilayers and heterostructures-far beyond the constraints of techniques such as terahertz emission, which are typically limited to ultrathin films and specific geometries. This methodology is particularly well suited to disentangle coexisting magnetization and spin transport dynamics in complex systems, enabling insights into superdiffusive spin transport, and offering a pathway to investigate for example ultrafast spin-charge interconversion, spin Seebeck effects, and magnetization control in emerging platforms including topological insulators, antiferromagnets, and two-dimensional materials.

Altogether, our findings provide a generalizable and experimentally accessible approach to track femtosecond spin currents and their influence on magnetization, opening new avenues for the design of ultrafast, energy-efficient spintronic devices unconstrained by material class or geometry.

*Acknowledgments*- The authors wish to thank Eric Fullerton for insightful exchanges. This work was supported by the France 2030 government investment plan managed by the French National Research Agency under grant references PEPR SPIN – TOAST ANR-22-EXSP-0003 and SPINMAT ANR-22- EXSP-0007. This work was also supported by the MAT-PULSE program, reference ANR-15-IDEX-04-LUE as part of Lorraine Université d'Excellence, an Excellence Initiative funded by France 2030. This work was also supported Project No. ANR-23-CE30-0047 SLAM., the Institute Carnot ICEEL, the Région Grand Est, the Metropole Grand Nancy for the project "OPTIMAG" and FASTNESS. J.I. also wishes to acknowledge the support of project JSPS KAKENHI JP 24K22964.